\documentclass[pra,12pt,a4paper]{revtex4}

\usepackage[cp866]{inputenc}
\usepackage[english]{babel}
\usepackage{amsmath}
\usepackage{amscd}
\usepackage{amssymb}
\usepackage{amsfonts}
\usepackage{dcolumn}
\usepackage{epsfig}

\newcommand{\inputEps}[3]{
            \centerline{\epsfxsize=#1\epsfbox{#3} }
            \vskip 2pt {\center\small{#2}} \vskip 2pt}

\newcommand{\Tr}{\operatorname{Tr}}

\begin{document}

\author{Fabio Benatti$^{a,b}$, Roberto Floreanini$^{b}$,
Sebastien Breteaux$^{c}$}
\affiliation{$^a$Dipartimento di Fisica Teorica, Universit\`a di Trieste,
Strada Costiera 11,\\
34014 Trieste, Italy\\
${}^b$Istituto Nazionale di Fisica Nucleare, Sezione di Trieste,
34100 Trieste, Italy\\
${}^c$Universit\'e de Rennes 1, 2 Rue du Thabor CS 46510, 35065
Rennes, France}

\title{\bf Slipped non-Positive Reduced Dynamics and Entanglement}

\begin{abstract}
Non-positive Markov approximations are sometimes used to describe
the dynamics of qubits in weak interaction with suitable
environments; the appearance of negative probabilities is avoided by
assuming that the transient regime eliminates from the possible
initial conditions those qubit states which would otherwise be
mapped out of the Bloch sphere by the subsequent Markovian
time-evolution. By means of a simple model, we discuss some physical
inconsistencies of this approach in relation to entanglement; in
particular, we show that slipped non-positive reduced dynamics might
create entanglement through a purely local action.
\end{abstract}

\maketitle

\section{Introduction}
\label{sec1}

Semigroups of dynamical maps are used to describe the time-evolution
of open quantum systems $S$ in weak interaction with suitable
external environments, typically an infinite heath bath in
equilibrium at a given temperature acting as a source of dissipation
and noise. They have been successfully used in many phenomenological
applications in quantum chemistry, quantum optics, statistical
physics \cite{Spo,AL,BP,BF0,BF1}.

In the following, we shall consider qubits described by
density matrices $\rho$, corresponding to three dimensional real
vectors of length $\leq 1$ in the Bloch sphere,
that evolve in time under the action of semigroups of linear maps
$\gamma_t$, $t\geq 0$.
Formally, these arise from the exponentiation of a
generator $\mathbb{L}$, $\gamma_t=\exp(t\mathbb{L})$, and satisfy
the forward in time composition law
$\gamma_t\circ\gamma_s=\gamma_{t+s}$, $t,s\geq 0$.

A preliminary natural request on the maps $\gamma_t$ is that they
send any initial state $\rho$ into another state $\gamma_t[\rho]$ at
all $t\geq 0$, namely that they map the Bloch sphere into itself.
Only in such a way, the spectrum of an evolving $\rho$ remains
positive and its eigenvalues can be interpreted as probabilities.

Such a property of the maps $\gamma_t$ is called
\textit{positivity}. In line of principle, it is not sufficient to
guarantee full physical consistency of the maps $\gamma_t$: a more
restrictive property, namely \textit{complete positivity}, need be
imposed \cite{CH1,Tak,Kra}. In such a way, not only the positivity
of $\gamma_t$ is guaranteed, but also that of the amplified map
$\gamma_t\otimes{\rm id}$; this describes the time-evolution of a
qubit $S$ statistically coupled to an ancillary qubit which remains
inert under the action of the identity operation ${\rm id}$.
Moreover, complete positivity of $\gamma_t$ fully characterizes the
form of the generator $\mathbb{L}$ \cite{GKS,Lin}. In particular, in
the case of a qubit system, there appears a characteristic order
relation ($2T_1\geq T_2$) between the decay times of the diagonal
($T_1$) and off-diagonal ($T_2$) entries of its time-evolving
density matrix \cite{GKS}.

The semigroup $\gamma_t$ describes the reduced dynamics of the
immersed qubit $S$ when the environment degrees of freedom have been
eliminated and the memory effects due to a short transient regime
have been got rid of by means of suitable Markovian approximations.
If not performed with due care, these latter
lead to reduced dynamics neither completely positive, nor even positive.

Absence of positivity of $\gamma_t$ means that there are initial
density matrices $\rho$ that may, in the course of time, develop
negative eigenvalues and thus lose their meaning as physical states.
If one wants to stick to non-positive reduced dynamics, a possible
way out of physical inconsistencies amounts to assuming that not all
density matrices are allowed as initial conditions for $\gamma_t$,
but only those which do not develop negative eigenvalues. The
mechanism which eliminates the unwanted initial states is ascribed
to the transient regime which rules the time behavior of the
subsystem $S$ before one can legitimately use the semigroup
$\gamma_t$. Namely, prior to $\gamma_t$, the environment action on
the subsystem is via a map $\mathbb{S}$ that projects the whole
state space of $S$, $\mathcal{S}(S)$, into a subset of ``good''
states $\mathbb{S}(\mathcal{S}(S))$, so that
$\gamma_t\circ\mathbb{S}$ acts as a positive map on
$\mathcal{S}(S)$, even if $\gamma_t$ does not.

The map $\mathbb{S}$ is known in the literature as \textit{slippage
of initial conditions} \cite{P,GH,SSO,GN,W}, its introduction being
motivated by the the difficulty to accept that a physical effect
like the decay-times hierarchy of a qubit be induced by its possible
entanglement with an uncontrollable external ancillary qubit, a
seemingly academic, abstract scenario with philosophical overtones.

On the other hand, quantum information and communication theories
have amply demonstrated the role of entanglement as a concrete
physical resource and developed techniques both theoretical and
experimental for its manipulation \cite{NC}. It is thus not only of
academical interest to study the slippage-approach in relation to
entanglement; more precisely, we shall be interested in setting the
ground to answer the following question. Suppose the inconsistencies
of a non-positive $\gamma_t$ have been cured by considering a
\textit{slipped} $\gamma_t\circ\mathbb{S}$. Which kind of effects
has $(\gamma_t\circ\mathbb{S})\otimes{\rm id}$ on the entangled
states of $S+S$?

Aim of this paper is to extend a previous result \cite{BFP} in order
to indicate which kind of pathologies may arise from amplifications
of slipped non-positive semigroups acting on bipartite systems.
Given an environment, its associated slippage operator $\mathbb{S}$
is far from being technically accessible as a mathematical object
and such are its effective properties. We shall thus model
$\mathbb{S}$ by the simplest map with the prescribed slippage
properties and then discuss its action on the isotropic states
\cite{HHH0} of a two qubit systems, one qubit immersed in a
dephasing environment and the other one external to it. We shall
show that, by preceding it by a suitable slippage operator
$\mathbb{S}$, a non-positive reduced dynamical map $\gamma_t$ can
indeed be turned into a completely positive map
$\gamma_t\circ\mathbb{S}$, but also that, unless strongly slipped,
the maps $\gamma_t$ can, acting locally, create entanglement, a
fundamentally non-local property. Furthermore, we shall present
instances of cases where the necessary strong slippage results in
the elimination of all isotropic entangled states. We take this as
an indication of a general behavior: in the weak-coupling regime,
slipped non-completely positive dynamics appear to be incompatible
with entanglement.

\section{Quantum Dynamical Semigroups}
\label{sec2}

In this section, we shall briefly review the standard approach to
open quantum dynamics, namely the so-called weak-coupling limit.

The typical physical context is represented by a finite-dimensional
($n$-level) subsystem $S$ in weak interaction with an environment
$E$, this latter corresponding to an infinite dimensional reservoir
in equilibrium at some fixed temperature $T$. The space of states
$\mathcal{S}(S)$ of $S$ consists of $n\times n$ density matrices
$\rho_S$, while the state of the environment $\rho_E$ is taken to be
an equilibrium state $\rho_E$. The dynamics of the (closed) system
$S+E$ is typically described by a Hamiltonian of the form
\begin{equation}
\label{Ham1} H_{S+E}=H_S+H_E+\,g\,  H_I\ ,\quad H_I=\sum_\alpha
X^\alpha_S\otimes X^\alpha_E\ ,
\end{equation}
where $H_S$ is the subsystem Hamiltonian, $H_I$ an interaction term,
of strength $g$, linear in the system and (centered) environment
operators $X^\alpha_{S,E}$ ($\Tr_E(\rho_EX^\alpha_E)=0$).

When the interaction between $S$ and $E$ is weak ($g<<1$), it makes
sense to derive a so-called \textit{reduced dynamics}, namely a
description of the time-evolution of $S$ involving the system $S$
alone. The derivation amounts to the elimination of the degrees of
freedom of $E$ while retaining the effects of their presence on $S$;
the resulting time-evolution is irreversible, in general non-linear,
and dominated by memory effects. By assuming the initial state of
$S+E$ to be of the uncorrelated form $\rho_S\otimes\rho_E$,
non-linearities are automatically eliminated and the standard
\textit{projection technique} leads to the following equation of
motion for $\rho_S\in\mathcal{S}(S)$ \cite{AL}:
\begin{equation}
\label{memoryeq1}
\partial_t\rho_{S}(t)=-i\biggl[H_S\,,\,\rho_S(t)\biggr]+\,g^2\,
\int_0^t{\rm d}s\, \mathbb{K}(s)[\rho_{S}(t-s)]\ ,
\end{equation}
where $\mathbb{K}(s)$ is a highly complicated kernel that, acting on
the states of $S$, takes into account both the degrees of freedom of
$E$ and the time-evolution of all of them prior to time $t$.

The ensuing dynamics preserves the trace of $\rho_S$ and is
completely positive in the sense that it sends any initial $\rho_S$
into a $\rho_S(t)$ which results from the action of a linear map
$\mathbb{G}_t$:
\begin{equation}
\label{memoryeq2}
\rho\mapsto\rho_{S}(t)=:\mathbb{G}_t[\rho_S]=\sum_j V_j(t)\,\rho_S\,
V^\dagger_j(t)\ ,
\end{equation}
where the $V_j(t)$ are $n\times n$ matrices such that
$\sum_jV_j^\dagger(t)V_j(t)=1$.
\bigskip

\noindent
{\bf Remarks 2.1}\hfill

\begin{enumerate}
\item
A trace-preserving linear map $\mathbb{G}$ on $\mathcal{S}(S)$ is
called \textit{positive} if $\mathbb{G}[\rho]$ has positive spectrum
for any $\rho\in\mathcal{S}(S)$. If $\mathbb{G}$ is
\textit{completely positive}, then it is not only positive, but,
upon coupling $S$ with another $n$-level system system, the
amplified map $\mathbb{G}\otimes{\rm id}$ preserves the positivity
of all possible states of the compound system $S+S$, where ${\rm
id}$ means that the ancillary system $S$ is not affected. It turns
out that complete positivity is equivalent to $\mathbb{G}$ being of
the Kraus-Stinespring form \cite{Tak,Kra}
\begin{equation}
\label{KStin}
\mathbb{G}[\rho]=\sum_jG_j\,\rho\,G_j^\dagger\ ,\quad
\sum_jG_j^\dagger G_j=1\ .
\end{equation}
\item
The physical meaning of complete positivity is intimately related to
the phenomenon of quantum entanglement \cite{BF1}; therefore, its
physical role can be appreciated only considering correlated
bipartite quantum systems of which only one party undergoes a state
change. In fact, if coupling of $S$ to ancillas could be excluded,
the physical transformations of the states of $S$ might adequately
be described by positive linear maps $\mathbb{G}$. Also, if $S$ is
coupled to another $S$, amplified positive maps
$\mathbb{G}\otimes{\rm id}$ would preserve the positivity of any
separable state
\begin{equation}
\label{sepstates}
\rho_{S+S}^{sep}:=\sum_{ij}\lambda_{ij}\rho_S^i\otimes\rho_S^j\
,\qquad\rho_S^{i,j}\in\mathcal{S}(S)\ .
\end{equation}
However, since there are states of $S+S$, the \textit{entangled}
ones, which cannot be written as in~(\ref{sepstates}), when
$\mathbb{G}\otimes{\rm id}$ acts on anyone of them, the positivity
of the eigenvalues of $\mathbb{G}\otimes{\rm id}[\rho^{ent}_{S+S}]$
can be preserved if and only if $\mathbb{G}$ is not only positive,
but also completely positive.
\item
If a linear map $\mathbb{G}$ on the states of $S$ is not completely
positive, this does not mean that $\mathbb{G}\otimes{\rm id}$
sends any entangled state of $S+S$ out of the class of states:
some entangled states may none the less have the positivity of their
spectrum preserved at all times.
However, for any non-completely positive $\mathbb{G}$,
the totally symmetric projection
\begin{equation}
\label{symproj} P:=\frac{1}{n}\sum_{i,j=1}^n\,|i\rangle\langle
j|\otimes|i\rangle\langle j|\ ,
\end{equation}
will always have some negative eigenvalues appearing in the spectrum
of $\gamma_t\otimes{\rm id}[P]$. Indeed, a theorem of Jamolkiowski
\cite{HHH0} establishes that $\mathbb{G}\otimes{\rm id}[P]$ is a
positive matrix if and only if $\mathbb{G}$ is completely positive.
\end{enumerate}
\bigskip

Beside being practically impossible to handle mathematically, the
dynamical maps $\mathbb{G}_t$ do not satisfy a forward-in-time
composition law: $\mathbb{G}_t\circ\mathbb{G}_s\neq
\mathbb{G}_{t+s}$. Manageable Markovian approximations are obtained
by weak-coupling limit techniques \cite{AL}; these exploit the
weakness of the interaction between $S$ and $E$.
Indeed,~(\ref{memoryeq1}) implies that the influence of the
environment becomes visible on a time scale such that $g^2t=\tau$;
by setting $t=\tau/g^2$ in~(\ref{memoryeq1}), the typical
prescription is to let $g\to0$ and approximate $\int_0^{\tau/g^2}$
by $\int_0^\infty$ and $\rho(t-s)=\rho(\tau/g^2-s)$ by
$\rho(\tau/g^2)=\rho(t)$. One thus obtains the equations
\begin{equation}
\label{memoryeq3}
\partial_t\rho_{S}(t)=-i\biggl[\widetilde{H}_S\,,\,\rho_S(t)\biggr]+
g^2\,\mathbb{D}[\rho_S(t)]\ ,
\end{equation}
where $\mathbb{D}:=\int_0^\infty{\rm d}s\,\mathbb{K}(s)$ and
$\widetilde{H}_S$ is a Lamb-shifted subsystem Hamiltonian corrected
by terms of order $g^2$.

The linear map $\mathbb{D}$ takes into account the environment
induced dissipative effects, essentially damping and noise,
affecting the subsystem $S$ when it weakly interacts with its
environment $E$. By introducing a Hilbert-Schmidt orthonormal set of
$n\times n$ matrices $F_i$, $i=1,2,\ldots, n^2-1$,
$F_{n^2}=1/\sqrt{n}$, $\Tr(F_i^\dagger F_j)=\delta_{ij}$,
$\mathbb{D}$ can always be put in the form
\begin{equation}
\label{memoryeq4} \mathbb{D}[\rho_S]=\sum_{i,j=1}^{n^2-1}
C_{ij}\Bigl(F_i\,\rho_S\,F^\dagger_j\,-\,
\frac{1}{2}\Bigl\{F^\dagger_jF_i\,,\,\rho\Bigr\}\Bigr)\ ,
\end{equation}
where $C=[C_{ij}]$ is a self-adjoint $(n^2-1)\times(n^2-1)$
coefficient matrix, the so-called \textit{Kossakowski matrix}
\cite{GKS}.

By setting the right hand side of~(\ref{memoryeq4}) equal to
$\mathbb{L}[\rho_S(t)]$, the resulting Markovian reduced dynamics
consists of a semigroup of linear maps $\gamma_t=\exp(t\mathbb{L})$;
these maps are trace-preserving and their positivity or complete positivity
depends on the matrix $C$:
\bigskip

\begin{enumerate}
\item
if the Kossakowski matrix $C\geq 0$, then the maps $\gamma_t$ are
completely positive and vice versa~\footnote{\label{foot1} The
derivation of the reduced dynamics as sketched in Section II does
not lead to either completely positive or positive $\gamma_t$, in
general. In order to get a semigroup of such maps, a sufficient
prescription is, roughly speaking, to formally
integrate~(\ref{memoryeq1}) and then operate an ergodic average of
the kernel (see \cite{AL} for more details)};
\item
if $C$ is not positive, no general necessary and sufficient
conditions are known which yield positive $\gamma_t$ (see \cite{BF1}
for certain cases where this possibility indeed exists).
\end{enumerate}
\bigskip

\section{An open qubit system}
\label{sec3}

We now follow and develop an argument developed in \cite{BFP} and
consider, as a concrete model, a single qubit, represented by a spin
$1/2$ particle coupled to a constant magnetic field vertically
directed and to a weak stochastic classical magnetic field
$\mathbf{G}=(G_1(t),G_2(t),G_3(t))$. The time-evolution of the qubit
density matrices is determined by the time-dependent Liouville-von
Neumann equation
\begin{equation}
\label{Ham2}
\partial_t\rho_S(t)=-i\biggl[\widetilde{\omega}\,\sigma_3\,,\,
\rho_S(t)\biggr]\,
-i\,\biggl[\sum_{i=1}^3\, G_i(t)\,\sigma_i\,,\,\rho_S(t)\biggr]\ ,
\end{equation}
where the $\sigma_i$ are the Pauli matrices and the scalar quantities
$G_i(t)$ are chosen to be stationary, stochastic variables
with vanishing first moments and diagonal covariance matrix:
\begin{equation}
\label{Stoc1}
\langle G_i(t)G_j(s)\rangle=G_i{\rm e}^{-\lambda_i(t-s)}\delta_{ij}\ ,
\end{equation}
with $G_1>G_2>0$, $G_3>0$ and  $\lambda_1=\lambda_2=\lambda>0$.
Such a classical stochastic field
is a convenient modeling of a heat bath whose temperature is
sufficiently high, whereby the global
Hamiltonian~(\ref{Ham1}) for the system plus environment may be replaced
by~(\ref{Ham2}).

Of course, the solution of the above stochastic equation is a
stochastic density matrix; useful physical information may only come
by considering the average of $\rho_S(t)$ with respect to the noise.
Techniques  based on the weak-coupling limit can be applied
to~(\ref{Ham2}) \cite{BS}; in general, these lead to a reduced
dynamics which suffers from lack of positivity of the ensuing
Markovian reduced dynamics.

With reference
to the evolution equation~(\ref{memoryeq3}) with the dissipative term
as in~(\ref{memoryeq4}), choosing
$\displaystyle F_i=\frac{\sigma_i}{\sqrt{2}}$, $i=1,2,3$, and
$\displaystyle F_4=\frac{1}{\sqrt{2}}$, one gets
\begin{eqnarray}
\label{Ham3}
&&\widetilde{H}=\omega\sigma_3\
,\quad\omega:=\widetilde{\omega}\biggl(1+
\frac{2(G_1+G_2)}{\lambda^2+4\widetilde{\omega}^2}\biggr)\\
\label{Ham4}
&& C=\begin{pmatrix}
\alpha_1&-b&0\cr-b&\alpha_2&0\cr0&0&a
\end{pmatrix}\ ,\quad
\alpha_{1,2}:=\frac{2G_{1,2}\lambda}{\lambda^2+4\,\widetilde{\omega}^2}
\ ,\ a=\frac{2G_3}{\lambda_3}\ , \
b:=2\widetilde{\omega}\frac{G_2-G_1}{\lambda^2+4\,\widetilde{\omega}^2}
\ .
\end{eqnarray}

In order to better expose the consequences of sticking to
non-positive dissipative semigroups, we simplify the analysis by
choosing $\widetilde{\omega}\ll\lambda_3\ll\lambda$ and
$\lambda\,G_{1,2}\ll\widetilde{\omega}\,|G_2-G_1|$. This allows us
to neglect $\alpha_{1,2}$ with respect to both $b$ and $a$; we shall
thus deal with the dissipative evolution equation
\begin{equation}
\label{Ham6}
\partial_t\rho_S(t)=-i\bigl[\omega\sigma_3\,,\,\rho_S(t)\bigr]\,+\,
a\biggl(\sigma_3\,\rho_S(t)\,\sigma_3\,-\,\rho_S(t)\biggr)\,-\,b\,
\biggl(\sigma_1\,\rho_S(t)\,\sigma_2\,+\,\sigma_2\,\rho_S(t)\,\sigma_1
\biggr)\ .
\end{equation}

In order to solve the above equation and study the ensuing
time-evolution, it proves convenient to recast the qubit density
matrices as
\begin{eqnarray}
\label{dens-mat1}
\rho=\begin{pmatrix}
\rho_{11}&\rho_{12}\cr\rho_{12}^*&1-\rho_{11}
\end{pmatrix}=\frac{1}{2}(1+\mathbf{r}\cdot\mathbf{\sigma})\ ,
\end{eqnarray}
where $\sigma=(\sigma_1,\sigma_2,\sigma_3)$ and
$\mathbf{r}=(r_1,r_2,r_3)$, with
\begin{eqnarray}
\label{dens-mat2}
r_1=2\Re(\rho_{12})\ ,\ r_2=-2\Im(\rho_{12})\ ,\
r_3=2\rho_{11}-1\ .
\end{eqnarray}
The matrix $\rho$ is positive if and only if its determinant
\begin{equation}
\label{det}
{\rm Det}(\rho)=\frac{1-\|\mathbf{r}\|^2}{4}\geq 0\ ,
\end{equation}
whence each state is identified by a vector
$\mathbf{r}\in\mathbb{R}^3$ of norm
$\|\mathbf{r}\|\leq1$, that is by a point in the so called Bloch
sphere.

If the qubit is weakly interacting with an environment,
its dissipative dynamics should be described by
a semigroup of trace-preserving maps $\gamma_t$ sending any state
$\rho$ at $t=0$ into a state $\gamma_t[\rho]$ at time $t>0$.
According to~(\ref{det}), this is equivalent to $|\mathbf{r}_t|\leq1$,
where $\mathbf{r}_t=(r_1(t),r_2(t),r_3(t))\in\mathbb{R}^3$
is the Bloch vector identifying
\begin{equation}
\label{deph1}
\gamma_t[\rho]=\frac{1}{2}\Bigl(1+\sum_{i=1}^3r_i\gamma_t[\sigma_i]\Bigr)
=\frac{1}{2}\Bigl(1+\sum_{i=1}^3r_i(t)\sigma_i\Bigr)\ .
\end{equation}

In particular,~(\ref{Ham6}) translates into a linear
differential equation for the Bloch vector
\begin{equation}
\label{deph2}
\begin{pmatrix}
\dot{r}_1\cr\dot{r}_2\cr\dot{r}_3
\end{pmatrix}=-2\mathcal{L}\begin{pmatrix}
r_1\cr r_2\cr r_3
\end{pmatrix}\ ,
\end{equation}
where the $3\times 3$ matrix $\mathcal{L}=\mathcal{H}+\mathcal{D}$
consists of the sum of an antisymmetric component $\mathcal{H}$
corresponding to the commutator with the Hamiltonian
in~(\ref{memoryeq3}), while $\mathcal{D}$ is symmetric and
corresponds to the dissipative term $\mathbb{D}$. Explicitly,
\begin{equation}
\label{deph3}
\mathcal{L}=\begin{pmatrix}
a& b+\omega& 0\cr
b-\omega&a&0\cr
0&0&0
\end{pmatrix}\ ,\ \mathcal{H}=\begin{pmatrix}
0&\omega& 0\cr
-\omega&0&0\cr
0&0&0
\end{pmatrix},\
\mathcal{D}=
\begin{pmatrix}
a& b& 0\cr
b&a&0\cr
0&0&0
\end{pmatrix}\ .
\end{equation}
The semigroup of maps $\gamma_t={\rm e}^{t\mathbb{L}}$ on the
state-space $\mathcal{S}(S)$ generated
by~(\ref{Ham6}) corresponds to a semigroup of $3\times 3$ matrices
$\mathcal{G}_t={\rm e}^{-2t\mathcal{L}}$ acting on the Bloch sphere.

From~(\ref{deph2}),~(\ref{deph3}) and~(\ref{Ham6}) it follows that
\begin{equation}
\label{deph4}
\left\{\begin{array}{l}
\dot{r}_1=-2a\,r_1-2(b+\omega)\,r_2\\
\dot{r}_2=-2(b-\omega)\,r_1-2a\,r_2\\
\dot{r}_3=0
\end{array}\right.\ .
\end{equation}
Setting $\Omega^2=\omega^2-b^2$ one gets:
\begin{equation}
\label{deph6a}
\left\{\begin{array}{l}
r_1(t)={\rm e}^{-2at}
\biggl[r_1\cos(2\Omega t)
-r_2\frac{\omega+b}{\Omega}\sin(2\Omega t)\biggr]\\
\\
r_2(t)={\rm e}^{-2at}\biggl[
r_1\frac{\omega-b}{\Omega}\sin(2\Omega t)+r_2\cos(2\Omega t)\biggr]\\
r_3(t)=r_3\hskip 8cm\hbox{and}
\end{array}\right.
\end{equation}

Apparently, the action of the semigroup is that of a dephasing channel
with no influence on the diagonal elements of any initial $\rho$: it
remains to be checked whether the matrices $\mathcal{G}_t$ correspond
to non-positive, positive or completely positive maps $\gamma_t$.

The Kossakowski matrix in~(\ref{Ham4}) is
positive, hence $\gamma_t$ completely positive, if and only if
$b=0$.

Let instead $b\neq0$ and consider the Bloch vectors
$\mathbf{r}_\pm=1/\sqrt{2}(0,1,\pm 1,0)$ corresponding to the pure
states (${\rm Det}(\rho_\pm)=0$)
$\rho_\pm=\frac{1}{2}(1+\sigma_1\pm\sigma_2)$. At $t>0$ they evolve
into matrices of trace $1$ with Bloch vectors such that
\begin{equation}
\label{deph7} \|\mathbf{r}_t^\pm\|^2={\rm e}^{-4at}\biggl[
\cos^2(2\Omega t)+\frac{\omega^2+b^2}{\Omega^2}\sin^2(2\Omega t)
\mp\frac{b}{\Omega}\sin(4\Omega t)\biggr]\simeq 1-4t(a\pm b)\ ,
\end{equation}
for $t\to0$. It thus follows that $\gamma_t$ acts in a physically
consistent way on the states of the qubit only if $a^2-b^2\geq0$.
Actually, this condition is also sufficient for $\gamma_t$ to
preserve the positivity of states for all $t>0$. Indeed,
$a^2-b^2\geq0$ corresponds to the positivity of the matrix
$\mathcal{D}$ in~(\ref{deph3}). Now, a state $\rho$ exits the Bloch
sphere  at time $t$ if and only if the Bloch vector of
$\gamma_t[\rho]$ is such that $\mathbf{r}_t=1$ and
\begin{equation}
\label{deph8}
\frac{{\rm d}\|\mathbf{r}_t\|^2}{{\rm d}t}
=-2\langle\mathbf{r}_t|\mathcal{D}|\mathbf{r}_t\rangle
=-2a(r_1^2(t)+r_2^2(t))-2br_1(t)r_2(t)>0\ .
\end{equation}
Thus, when $\mathcal{D}\geq 0$, the state moves from the surface
towards the interior of the Bloch sphere if
$\langle\mathbf{r}_t|\mathcal{D}|\mathbf{r}_t\rangle>0$, while if
$\langle\mathbf{r}_t|\mathcal{D}|\mathbf{r}_t\rangle=0$, then it
remains of norm $1$.
\bigskip

\noindent \textbf{Proposition 3.1}\quad The semigroup
$\mathcal{G}_t={\rm e}^{-2t\mathcal{L}}$ generated by~(\ref{deph2})
with $\mathcal{L}$ as in~(\ref{deph3}) corresponds to a semigroup of
linear maps $\gamma_t$ on the state-space of the qubit $S$ which are

\begin{enumerate}
\item
completely positive if and only if $b=0$;
\item
only positive if and only if $b\neq0$ and $a^2-b^2\geq 0$;
\item
not even positive if $0\leq a<b$.
\end{enumerate}

\section{Slippage of initial conditions}
\label{sec4}

From the considerations in Remark 2.1.2, in order to avoid
physical inconsistencies that may arise from non-completely positive
reduced dynamics in presence of entangled states, it would be rather natural
to consider only reduced dynamics enjoying complete positivity and to
exclude those that do not.
Actually, the problem is more acute since those Markovian
approximations $\gamma_t$ which fail to be completely positive are
sometimes not even positive.
This means that their pathological behaviour, of the kind discussed in
the previous section, already appears with a single qubit.

At the level of a single system, if one wants to stick to
non-positive reduced dynamics as the best Markovian approximations
to the actual dissipative dynamics, a possible way out of the
physical inconsistencies is known in the literaure as
\textit{slippage of initial conditions} \cite{P,GH,SSO,GN,W}.

The argument goes roughly as follows. Not all density matrices
$\rho_S$ evolve into $\gamma_t[\rho_S]$ with negative eigenvalues in
their spectra; thus, let $\mathcal{S}^*(S)\subseteq\mathcal{S}(S)$
denote the subset of states which remain positive under $\gamma_t$
for all $t>0$ and let
$\mathbb{S}:\mathcal{S}(S)\mapsto\mathcal{S}^*(S)$ be a map
selecting this subset of ``good states'' for $\gamma_t$. Then, even
though the maps $\gamma_t$ suffer from physical inconsistencies, the
maps $\gamma_t\circ\mathbb{S}$ do not.

The slippage of initial conditions is supposed to be
operated by the transient regime prior to the
Markov one. In other words, according
to~(\ref{memoryeq1}),
\begin{equation}
\mathbb{S}[\rho_S]\simeq\,g^2\,\int_0^{t_{trans}}{\rm d}s\,
\mathbb{K}(s)[\rho_S(t_{trans}-s)]\ ,
\end{equation}
where $t_{trans}$ is the span of time during which memory effects
cannot be disregarded. Notice that the weak-coupling limit $g^2
t=\tau$, $g\to0$, does exactly push $t_{trans}\to 0$. Roughly
speaking, the action of the transient regime is supposed to be such
that the Markovian dynamics following it does not act on any
possible initial density matrix, but only on those selected by
$\mathbb{S}$ which are free from the inconsistencies arising from
non-positivity.

Of course the plausibility of such an approach has to be tested
against actual physical open quantum contexts; however, the highly
complex nature of the kernel $\mathbb{K}(t)$ describing the transient,
memory-full dynamics, scarcely allows any
hope to go beyond extremely rude approximations in describing
mathematically what its effects really are.
In the following, relative to the reduced dynamics in~(\ref{deph6a}),
we shall model a slippage operator in a simple way
that however embodies some of its salient features.

We first observe that according to the slippage philosophy, there
would be no need to invoke a selection map $\mathbb{S}$
when $a^2\geq b^2$ for in this case the maps
$\gamma_t$ are positive and cannot drive any initial state outside the
Bloch sphere.
In other words, all initial $2\times 2$ density matrices
are good initial states for the reduced dynamics and one need not
resort to a slippage mechanism to select a suitable subclass of them.

We shall then set $a^2<b^2$;
from the arguments of the previous section, the
most dangerous states are pure, so that non-pure states within the
Bloch sphere are expected to offer less problems.
How far into the Bloch sphere should one go in order to be sure to
get rid of all those states which $\gamma_t$ would sooner or later map
outside it?
It is evident from~(\ref{deph6a}) that Bloch vectors $\mathbf{r}_t$
may exit the Bloch sphere but their lengths $\|\mathbf{r}_t\|$
cannot diverge.
The maximum value $R$ achievable by $\|\mathbf{r}_t\|$ can be
explicitly obtained.

First, observe that, within and on the Bloch sphere,
\begin{equation}
\label{slip00}
\|\mathbf{r}_t\|^2=
\langle\mathbf{r}|\mathcal{G}_t^T\mathcal{G}_t|\mathbf{r}\rangle
\leq\|\mathcal{G}_t^T\mathcal{G}_t\|\|\mathbf{r}\|^2
\leq\|\mathcal{G}_t^T\mathcal{G}_t\|\ ,
\end{equation}
where $\mathcal{G}_t^T$ is the transposed of $\mathcal{G}_t$ and the operator
norm $\|\mathcal{G}_t^T\mathcal{G}_t\|$ equals the largest eigenvalue
of the positive, real symmetric matrix
\begin{eqnarray}
\nonumber
\mathcal{G}_t^T\mathcal{G}_t&=&\begin{pmatrix}
0&0&0\cr0&0&0\cr0&0&1
\end{pmatrix}+{\rm e}^{-4at}\begin{pmatrix}
1&0&0\cr0&1&0\cr0&0&0\end{pmatrix}\\
\label{slip01} &+&{\rm e}^{-4at}\frac{2b}{\Omega^2}
\sin(2t\Omega)\begin{pmatrix}
(b-\omega)\,\sin(2t\Omega)&-\Omega\,\cos(2t\Omega)&0\cr
-\Omega\,\cos(2t\Omega)&(b+\omega)\,\sin(2t\Omega)&0\cr0&0&0
\end{pmatrix}
\ .
\end{eqnarray}

Of the three eigenvalues of the above matrix, one is $1$ and the
largest of the remaining two is
\begin{equation}
\label{slip02} R^2(t):={\rm
e}^{-4at}\Biggl(\frac{b}{\Omega}\left|\sin(2t\Omega)\right|+
\sqrt{1+\frac{b^2}{\Omega^2}\sin^2(2t\Omega)}\Biggr)^2\ .
\end{equation}
If $a^2<b^2$, it can be explicitly computed that $R(t)$ achieves its
maximum
\begin{equation}
\label{slip03} R:={\rm e}^{-2at^\prime}
\sqrt{\frac{\omega+\sqrt{b^2-a^2}}{\omega-\sqrt{b^2-a^2}}}
\qquad\hbox{at} \quad
t^\prime:=\frac{1}{2\Omega}\arcsin\left(\frac{\Omega}{b}
\sqrt{\frac{b^2-a^2}{\Omega^2+a^2}}\right)\ .
\end{equation}
\bigskip

\noindent
{\bf Remarks 4.1}\hfill

\begin{enumerate}
\item
$R$ is surely $>1$; indeed, for $a^2<b^2$, $\gamma_t$ is not
positive, whence the norm of $\mathcal{G}_t^T\mathcal{G}_t$ cannot
be $\leq 1$.
\item
The maximum~(\ref{slip03}) is achieved at $t^\prime$ starting from
the pure state $\rho^*$ whose Bloch vector $\mathbf{r}^*$ is the
eigenvector of $\mathcal{G}_t^T\mathcal{G}_t$ relative to the
eigenvalue $R^2$.
\end{enumerate}
\bigskip

Set $0\leq\mu\leq R^{-1}$ and consider the following linear map on
$\mathcal{S}(S)$:
\begin{equation}
\label{slip2}
\rho\mapsto\mathbb{S}_\mu[\rho]
=\frac{1}{2}\bigl(1+\mu\mathbf{r}\cdot\sigma\bigr)
\end{equation}
Then, because of linearity,
\begin{equation}
\label{slip3}
\gamma_t\circ\mathbb{S}_\mu[\rho]=
\frac{1}{2}\bigl(1+\mu\mathbf{r}_t\cdot\sigma\bigr)\ ,\quad
\hbox{and}\quad \|\mu\mathbf{r}_t\|\leq\mu R\leq1\ ,
\end{equation}
whence no state can be mapped out of the Bloch sphere by
$\gamma_t\circ\mathbb{S}_\mu$ at any $t\geq 0$.

The map $\mathbb{S}_\mu$ is completely positive, its Kraus-Stinespring
form being
\begin{equation}
\label{slip4} \mathbb{S}_\mu[\rho]
=\frac{1+3\mu}{4}\,\rho\,+\,\frac{1-\mu}{4}\sum_{i=1}^3\sigma_i\,\rho\,\sigma_i\
.
\end{equation}
\bigskip

\noindent \textbf{Remark 4.2}\quad If the interaction of the
subsystem with the environment in equilibrium is weak, then the
hypothesis of an initial factorized state $\rho_S\otimes\rho_E$ is
not an unphysical restriction, as well as the hypotheisis that the
transient act as a linear completely positive operator.

The action of $\mathbb{S}_\mu$ is rather drastic in that it rigidly
maps the unit Bloch sphere onto a sphere of smaller radius; one
could instead envisage more sophisticated slippage mechanisms.
However, the intent of this paper is to show the kind of problems
afflicting the simplest possible of them. They should indeed not
depend on the explicit model of transient, being it related to the
non-positivity of the Markovian approximation and, as we shall
presently see, to the existence of entangled states.

\section{Slippage vs Entanglement}
\label{sec5}

Thanks to the rapid development of quantum information, the notion
of entanglement left the purely epistemological arena in which it
was confined for thirty or so years and made its appearance on the
foreground of physics as a most precious resource to perform
otherwise impossible tasks as dense quantum coding and quantum
teleportation \cite{NC}.

Physically speaking, the model we are going to consider is that of
one qubit immersed in a stochastic environment whose effects are
described by the maps $\gamma_t$ of the previous section, which is
correlated to another qubit which is a dynamically independent and
inert ancilla. This is the physical context when, for instance, an
entangled Bell state is constructed in a laboratory and, while one
of the two qubit is kept there, the other is sent to a distant party
through a noisy channel. The time-evolution of the compound system
is thus described by the semigroup of maps $\gamma_t\otimes{\rm
id}$, where the presence of the bath is felt locally by one of the
two qubits only and the other does not evolve at all (this is the
meaning of ${\rm id}$).

In the following, we shall study how the entanglement content of
special qubit states changes under the local time-evolution as
above. Concretely, we shall consider the time-evolution of the
\textit{concurrence} \cite{Woo}, which, for any state $\rho$ of a
two qubit system, is defined by
\begin{equation}
\label{ent3}
\mathcal{C}(\rho):=\max\{0,\lambda_1-\lambda_2-\lambda_3-\lambda_4\}\
,
\end{equation}
where the $\lambda_i$'s are the positive square roots of the
(positive) eigenvalues of the matrix
\begin{equation}
\label{ent4}
\rho\,\widetilde{\rho}\ ,\quad\widetilde{\rho}:=
(\sigma_2\otimes\sigma_2)\,\rho^*\,(\sigma_2\otimes\sigma_2)\ ,
\end{equation}
where $\rho^*$ is the matrix obtained by taking the complex
conjugate of the entries of $\rho$ in a chosen representation.

By means of the techniques sketched above, we shall now study whether
curing the non-positivity of reduced dynamics by slipping the initial
conditions can nevertheless conflict with the existence of
entanglement between the immersed open qubit and its inert ancilla.
\bigskip

\noindent \textbf{Remark 5.1}\quad That problems may easily arise
can be argued by observing that in the case of $a^2>b^2>0$, the maps
$\gamma_t$ are positive and, a priori, no slippage is needed.
Nevertheless, according to Remark 2.1.1, the totally symmetric
projector $P$ is definitely mapped out of the state of space of the
compound system $S+S$, for $\gamma_t$ is completely positive only if
$b=0$. Therefore, if $b\neq0$, one needs a slippage operator act on
the two qubit system; as the latter can only be due to the bath, it
must also affect the single qubit immersed in it, independently of
whether it is coupled or not to an ancillary qubit.
\bigskip

In order to set the ground for discussing the slippage mechanism in
relation to entanglement, we rewrite
the totally symmetric projector as follows
\begin{equation}
\label{slippent0}
P=\frac{1}{4}\Bigl(1\otimes
1+\sigma_1\otimes\sigma_1-\sigma_2\otimes\sigma_2
+\sigma_3\otimes\sigma_3\Bigr)\ ,
\end{equation}
so that~(\ref{slip4}) maps it into
\begin{equation}
\label{slippent1} P_\mu:=\mathbb{S}_\mu\circ{\rm
id}[P]=\frac{1}{4}\biggl(1\otimes
1+\mu\Bigl(\sigma_1\otimes\sigma_1-\sigma_2\otimes\sigma_2
+\sigma_3\otimes\sigma_3\Bigr)\biggr)\ .
\end{equation}

Then we let $\gamma_t\otimes{\rm id}$ act on $P_\mu$ and consider
$P_\mu(t):=\gamma_t\otimes{\rm id}[P_\mu]$; the result is
easily computed from
\begin{equation}
\label{deph6b}
\left\{\begin{array}{l}
\sigma_1(t)={\rm e}^{-2at}
\biggl[\sigma_1\cos(2\Omega t)+
\sigma_2\frac{\omega-b}{\Omega}\sin(2\Omega t)\biggr]\\
\sigma_2(t)={\rm e}^{-2at}\biggl[
-\sigma_1\frac{\omega+b}{\Omega}\sin(2\Omega t)
+\sigma_2\cos(2\Omega t)\biggr]\\
\sigma_3(t)=\sigma_3\ ,
\end{array}\right.
\end{equation}
which follow using~(\ref{deph6a}) and~(\ref{deph1}).
Explicitly,
\begin{equation}
\label{slippent2}
P_\mu(t)=\frac{1}{4}
\begin{pmatrix}
1+\mu&0&0&2\mu\,B_t\cr
0&1-\mu&2\mu\,C_t&0\cr
0&2\mu\,C^*_t&1-\mu&0\cr
2\mu\,B^*_t&0&0&1+\mu
\end{pmatrix}\ ,\
\left\{
\begin{array}{l}
B_t:={\rm e}^{-2at}\Bigl(\cos(2\Omega t)-i\frac{\omega}{\Omega}\sin(2\Omega
t)\Bigr)\cr
\cr
C_t:=i\,\frac{b}{\Omega}\,{\rm e}^{-2at}\sin(2\Omega t)
\end{array}\right.\ .
\end{equation}
The eigenvalues of this $4\times 4$ matrix are, in decreasing order,
\begin{eqnarray}
\label{eigv1} e^\mu_1(t)&:=&\frac{1}{4}\Biggl[1+\mu\Biggl(1+2{\rm
e}^{-2at}
\sqrt{1+\frac{b^2}{\Omega^2}\sin^2(2\Omega t)}\Biggr)\Biggr]\\
\label{eigv2} e^\mu_2(t)&:=&\frac{1}{4}\Biggl[1+\mu\Biggl(1-2{\rm
e}^{-2at}
\sqrt{1+\frac{b^2}{\Omega^2}\sin^2(2\Omega t)}\Biggr)\Biggr]\\
\label{eigv3} e^\mu_3(t)&:=&\frac{1}{4}\biggl[1-\mu\biggl(1-2{\rm
e}^{-2at}
\frac{b}{\Omega}\sin(2\Omega t)\biggr)\biggr]\\
\label{eigv4} e^\mu_4(t)&:=&\frac{1}{4}\biggl[1-\mu\biggl(1+2{\rm
e}^{-2at} \frac{b}{\Omega}\sin(2\Omega t)\biggr)\biggr]\ .
\end{eqnarray}

As already stressed, physical consistency asks that the slippage
operator $\mathbb{S}_\mu$ cure not only the non-positivity of
$\gamma_t$, but also ensure the positivity of the maps
$(\gamma_t\circ\mathbb{S}_\mu)\otimes{\rm id}$ acting on the the
entangled states of the compound system $S+S$. In particular,
$P_\mu(t)$ must correspond to a density matrix at all times $t\geq
0$.
\bigskip

\noindent
\textbf{Proposition 5.2}\quad
Set
\begin{equation}
\label{eigv5}
R_4(t):=1+2{\rm e}^{-2at}\frac{b}{\Omega}\sin(2\Omega t)\ ,
\end{equation}
the eigenvalues of $P_\mu(t)$ are positive if and only if
\begin{equation}
\label{eigv5a}
0\leq \mu\leq \frac{1}{R_4}\ ,\quad\hbox{where}\quad
R_4:=1+2{\rm e}^{-2at_*}\frac{b}{\sqrt{\Omega^2+a^2}}
\end{equation}
is the maximum of $R_4(t)$ which is achieved at
$t_*=\frac{1}{2\Omega}\arcsin{\frac{\Omega}{\sqrt{\Omega^2+a^2}}}$.
\bigskip

\noindent \textbf{Proof:}\quad Since the eigenvalues
$e^\mu_{3,4}(t)$ interchange when $\sin(2\Omega t)$ changes sign,
one can always consider $0\leq t\leq \frac{\pi}{4\Omega}$. Then, the
positivity of $e^\mu_4(t)$ requires $\mu\,R_4(t)\leq 1$ for all
$t\geq 0$. Vice versa, suppose $\mu\leq R_4^{-1}(t)$ for all $t\geq
0$ and denote by
\begin{equation}
\label{eigv5b}
R_1(t):=1+2{\rm e}^{-2at}\sqrt{1+\frac{b^2}{\Omega^2}\sin^2(2\Omega t)}
\end{equation}
the term multiplying $\mu$ in the largest eigenvalue $e^\mu_1(t)$.
Since $0\leq\mu\leq1$ and $b>0$, one estimates:
$\mu\,R_1(t)\leq 2+\mu\,R_4(t)\leq 3$,
whence $1\geq e_1^\mu(t)\geq e^\mu_2(t)\geq e_3^\mu(t)\geq
e_4^\mu(t)\geq 0$.\hfill$\blacksquare$
\medskip

In figures 1--3, the red line shows the time behaviour of the
largest eigenvalue $e^\mu_1(t)$ and the blue line the lowest
eigenvalue $e^\mu_4(t)$, for given values of $a$ and $b$, with
$\mu=1$, $\mu>R_4^{-1}$ and $\mu<R_4^{-1}$. Both functions are
computed with rescaled parameters $a\to a/\omega$, $b\to b/\omega$
and $t\to\omega t$; in this way $\Omega^2=1-b^2$ with $0\leq a,b\leq
1$. The green line corresponding to height $1$ is showed for the
sake of comparison with $e^\mu_1(t)$.
\bigskip

\inputEps{7cm}{Figure 1: $a=0.1$, $b=0.9$, $R_4^{-1}=0.25$, $\mu=1$.
Red line: $e^\mu_1(t)$. Blue line: $e^\mu_4(t)$} {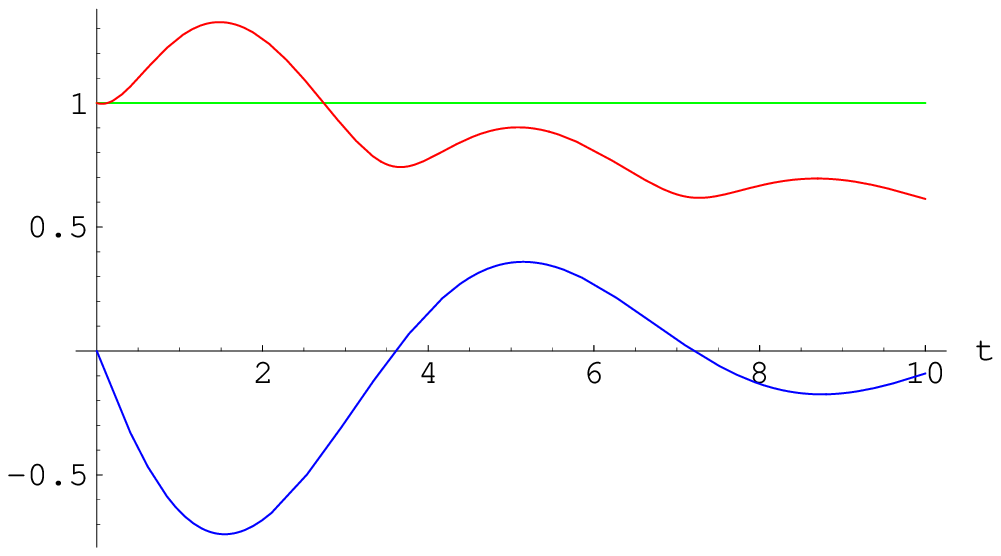}
\inputEps{7cm}{Figure 2: $a=0.1$, $b=0.9$, $R_4^{-1}=0.25$, $\mu=0.4$.
Red line: $e^\mu_1(t)$. Blue line: $e^\mu_4(t)$} {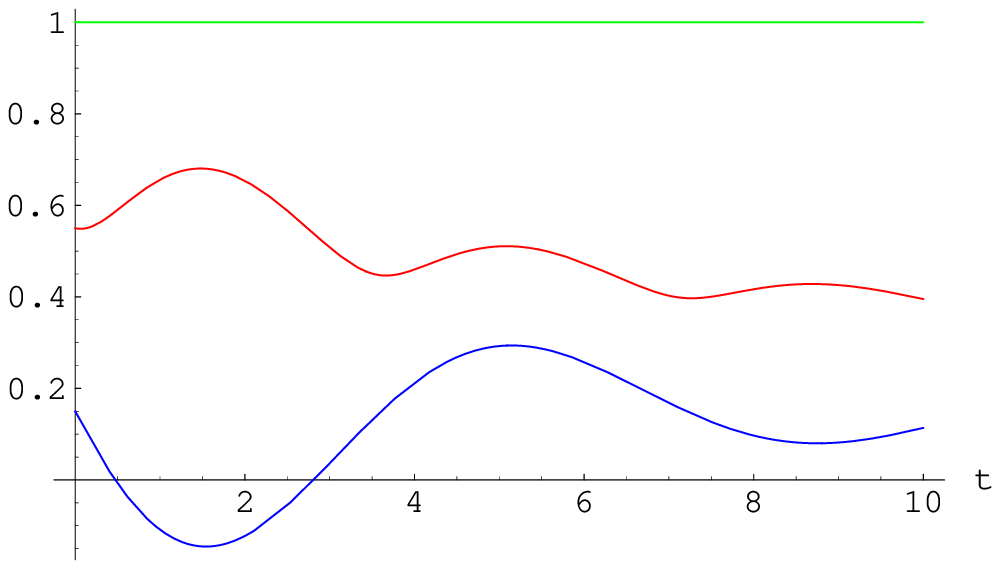}
\inputEps{7cm}{Figure 3: $a=0.1$, $b=0.9$, $R_4^{-1}=0.25$, $\mu=0.2$.
Red line: $e^\mu_1(t)$. Blue line: $e^\mu_4(t)$} {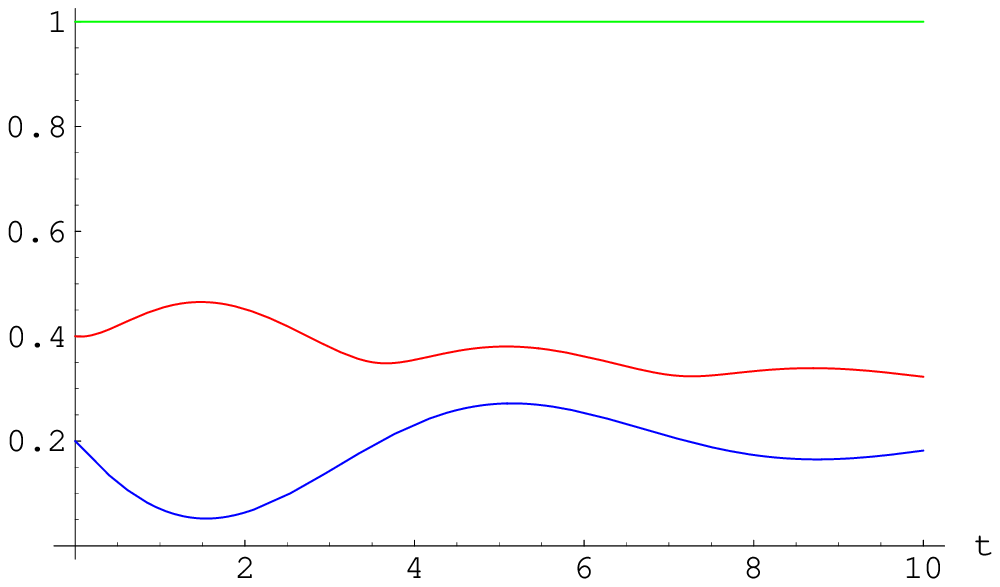} \vskip
2cm

\noindent
\textbf{Remarks 5.2}\hfill

\begin{enumerate}
\item
The states $P_\mu$ that are obtained from the totally symmetric
projector $P$ by the action of the slippage operator
$\mathbb{S}_\mu$ are the class of \textit{isotropic states},
$P_\mu=\frac{1-\mu}{4}+\mu\,P$ in standard form. They are entangled
if and only if $1\geq \mu>\frac{1}{3}$ \cite{HHH0}.
\item
According to Remark 2.1.2, condition~(\ref{eigv5a}) makes
$P_\mu(t)=(\gamma_t\circ\mathbb{S}_\mu)\otimes{\rm id}[P]$ a
positive matrix for all $t\geq 0$, whence all slipped maps
$\gamma_t\circ\mathbb{S}_\mu$ result completely positive, despite
$\gamma_t$ being possibly not even positive.
\item
If $a^2<b^2$, we have already seen that the channel $\mathbb{S}_\mu$
can cure non-positivity by multiplying the Pauli matrices by
$\mu\leq R^{-1}$ where $R$ is given in~(\ref{slip03}).
From~(\ref{slip02}) and~(\ref{eigv5}) it follows that $R(t)\leq
R_4(t)$ so that $R\leq R_4$; namely, the maximal radius of the
sphere of slipped initial conditions suggested by the single qubit
non-positive dynamics is too large for keeping physical consistency
against entanglement. The radius is to be decreased to at least
$R_4^{-1}$.
\end{enumerate}
\bigskip

In the following we shall see that even the value $R_4^{-1}$ does
not keep the non-positive dynamics free from pathologies; however,
the ones we are going to expose are much more intriguing and subtler
than the appearance of negative eigenvalues, being instead related
to the creation of entanglement by means of the local action of maps
of the form $\gamma_t\circ\mathbb{S}_\mu$.

We shall now consider $\mu\leq R_4^{-1}$ which ensures that the
$P_\mu(t)$ are physical states of the compound system $S+S$ at all
times $t\geq0$. In order to study their entanglement content, we
construct their concurrence $\mathcal{C}_\mu(t)$
(see~(\ref{ent3})--(\ref{ent4})); one checks that
$\widetilde{P}_\mu(t)=P_\mu(t)$, whence the square roots of the
eigenvalues of $P_\mu(t)\widetilde{P}_\mu(t)$ are the
values~(\ref{eigv1})--(\ref{eigv4}) and
\begin{equation}
\label{conc1} \mathcal{C}_\mu(t)=\max\{0,c_\mu(t)\}\ ,\qquad
c_\mu(t):=\mu\,{\rm e}^{-2at}\,
\sqrt{1+\frac{b^2}{\Omega^2}\sin^2(2\Omega t)}\,-\,\frac{1-\mu}{2}\
.
\end{equation}
At $t=0$, $\displaystyle c_\mu:=\frac{3\mu-1}{2}>0$ if and only if $\mu>1/3$,
that is if and only if $P_\mu$ is entangled.
\bigskip

\noindent \textbf{Proposition 5.3}\quad The states $P_\mu(t)$ are
entangled if and only if
\begin{equation}
\label{conc2} \frac{1}{R_1(t)}<\mu\leq\frac{1}{R_4}\ .
\end{equation}
\bigskip

\noindent \textbf{Proof:}\quad The upper bound on $\mu$ guarantees
that the $P_\mu(t)$ are well defined density matrices of the
bipartite system $S+S$ at all $t\geq 0$, while the lower bound
coming from~(\ref{conc2}), sets the range of $t\geq 0$ for which
$\mathcal{C}_\mu(t)=c_\mu(t)>0$. \hfill$\blacksquare$
\bigskip

\noindent \textbf{Remark 5.3}\quad Notice that by partially
transposing $P_\mu(t)$ in~(\ref{slippent2}), one exchanges $B_t$ and
$C_t$ so that the eigenvalues of $T\otimes{\rm id}[P_\mu(t)]$, where
$T$ denotes transposition, are the same of those $P_\mu(t)$, but
with $\mu\to-\mu$. Therefore, one checks that~(\ref{conc2})
corresponds to a non-positive partial transposed of $P_\mu(t)$, thus
to an entangled $P_\mu(t)$, while $\mathcal{C}_\mu(t)$ quantifies
the amount of entanglement it possesses.
\bigskip

We are now interested in the change of entanglement with time; from
Proposition 5.1, we know $\mathcal{C}_\mu(t)$ cannot increase if
$\gamma_t$ is completely positive. Indeed,
$\gamma_t\circ\mathbb{S}_\mu$, the composition of two completely
positive maps, would also be completely positive and hence a
physically consistent local operation.

Let us consider the time-derivative
\begin{equation}
\label{conc3}
\dot{c}_\mu(t)=\frac{2\mu\,{\rm e}^{-2at}}
{\sqrt{1+\frac{b^2}{\Omega^2}\sin^2(2\Omega t)}}\,G(t)\ ,\quad
G(t):=\biggl[\frac{b^2\sqrt{\Omega^2+a^2}}{\Omega^2}\cos(2\Omega t+\varphi)
\sin(2\Omega t)\,-\,a\biggr]\ ,
\end{equation}
where $\cos\varphi=\frac{\Omega}{\sqrt{\Omega^2+a^2}}$.
One checks that the function within square
brackets achieves its maximum
\begin{equation}
\label{conc4}
G:=\max_{t\geq0}G(t)=\frac{b^2}{2\Omega^2}\Bigl(\sqrt{\Omega^2+a^2}\,-\,a
\Bigr)-a\ ,
\end{equation}
at $\overline{t}=t_*/2$.

Further, $G>0$ if and only if $\displaystyle
a^2<\frac{b^4}{4\omega^2}$. This is only possible if $a^2<b^2$ for
$\omega^2>b^2$, hence only if $\gamma_t$ is non-positive; in such a
case $c_\mu(t)$ increases in a neighborhood of $\overline{t}$ and
does it independently of the value of $\mu$. Therefore, if in the
same neighborhood $\mathcal{C}_\mu(t)>0$, then the entanglement of
$P_\mu(t)$ increases by the local action of the non positive
$\gamma_t$, an apparent physical inconsistency, for, as already
noticed, physically, entanglement can be created by non-local
operations only.

In order to check this possibility, we compare the bound
$\mu\,R_4\leq 1$, which is necessary to physical consistency of the
$P_\mu(t)$ as states at all times, with the lower bound which
ensures positive concurrence. Since we are interested in a
neighborhood of $\overline{t}$, we shall consider
$0\leq\overline{t}+t$ as temporal parameter; then~(\ref{conc2})
gives
\begin{equation}
\label{conc5} R_1(\overline{t}+t)>R_4\Leftrightarrow f(t):={\rm
e}^{-2at}\,
\sqrt{1+\frac{b^2}{\Omega^2}\sin^2(2\Omega(\overline{t}+t))} \,-\,
{\rm e}^{-2a\overline{t}}\frac{b}{\sqrt{\Omega^2+a^2}}\,>\,0\ .
\end{equation}

In Figures 4--7, with the parameters $a,b,\omega$ rescaled as in
figures 1--3, the red lines show where $f(t)>0$, the blue ones where
$g(t):=G(\overline{t}+t)>0$ and thus the derivative in~(\ref{conc3})
is positive. It is then apparent that there are values of $a,b$ with
$a<b^2/(2\omega)$ ($a<b^2/2$ in the rescaled parameters) such that
there exist time-intervals $t\in[t_1,t_2]$ where~(\ref{conc5}) holds
together with $\dot{c}_\mu(\overline{t}+t)>0$. For such choices of
$a$ and $b$, in order to avoid the unphysical creation of
entanglement by means of local operations, one must exclude
$\mu\in[R_1(\overline{t}+t)^{-1},R_4^{-1}]$ with $t\in[t_1,t_2]$ and
must thus enforce the stronger bound
\begin{equation}
\label{conc6}
0\leq\mu\leq\frac{1}{\max_{t\in[t_1,t_2]}R_1(\overline{t}+t)}\ .
\end{equation}
Furthermore, in the same figures 4--7, the green lines plot the
function $r(t):=R_1(\overline{t}+t)-3$; then, figures 5 and 7 show
that, in the interval $[t_1,t_2]$ where $f(t)>0$ and $g(t)>0$, by
decreasing $a$, it also holds $r(t)\geq0$, whence that
$R_1(\overline{t}+t)^{-1}\leq1/3$ and, consequently, also the upper
bound in~(\ref{conc6}) is samller then $1/3$. According to Remark
5.2.1, it turns out that, in such cases, in order to ensure physical
consistency, $\mathbb{S}_\mu$ must destroy all entangled isotropic
states.

\inputEps{7cm}{Figure 4: $a=0.3$, $b=0.8$. Red line: $f(t)$. Blue line: $g(t)$.
Green line: $r(t)$}{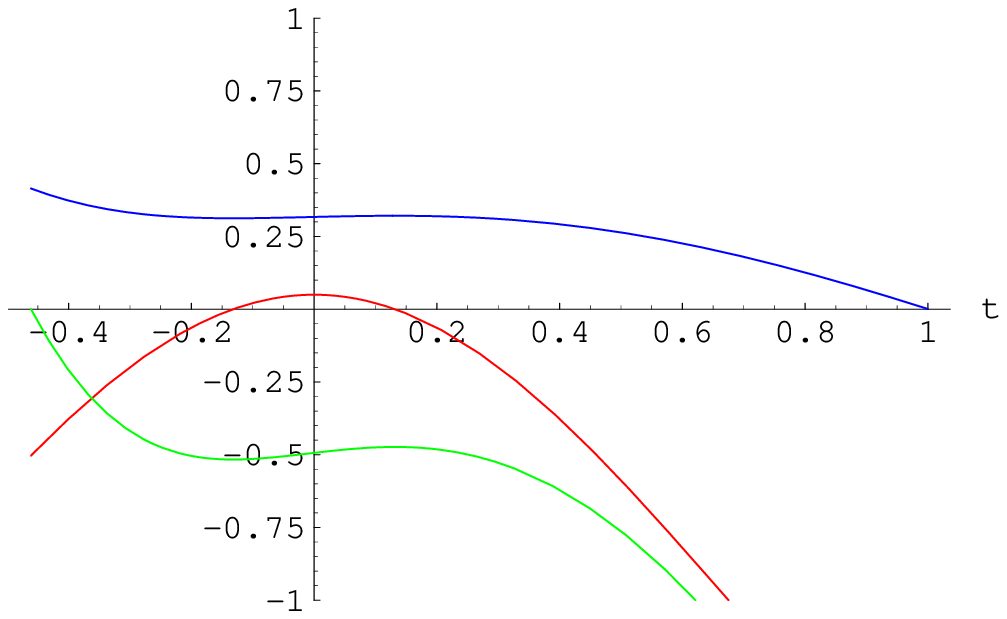}
\inputEps{7cm}{Figure 5: $a=0.1$, $b=0.8$. Red line: $f(t)$. Blue line: $g(t)$.
Green line: $r(t)$}{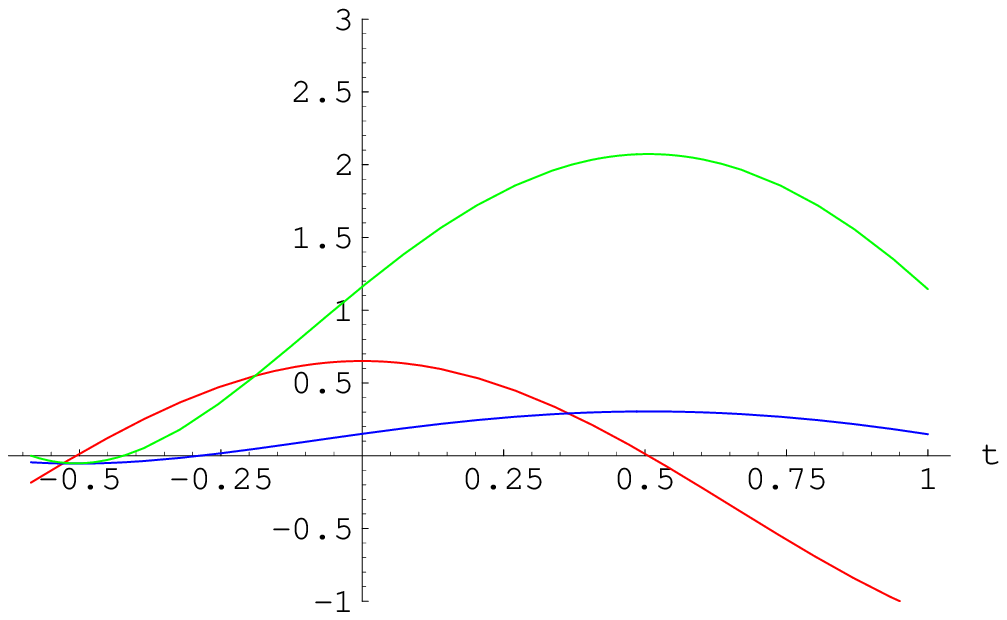}
\inputEps{7cm}{Figure 6: $a=0.06$, $b=0.4$. Red line: $f(t)$. Blue line: $g(t)$.
Green line: $r(t)$}{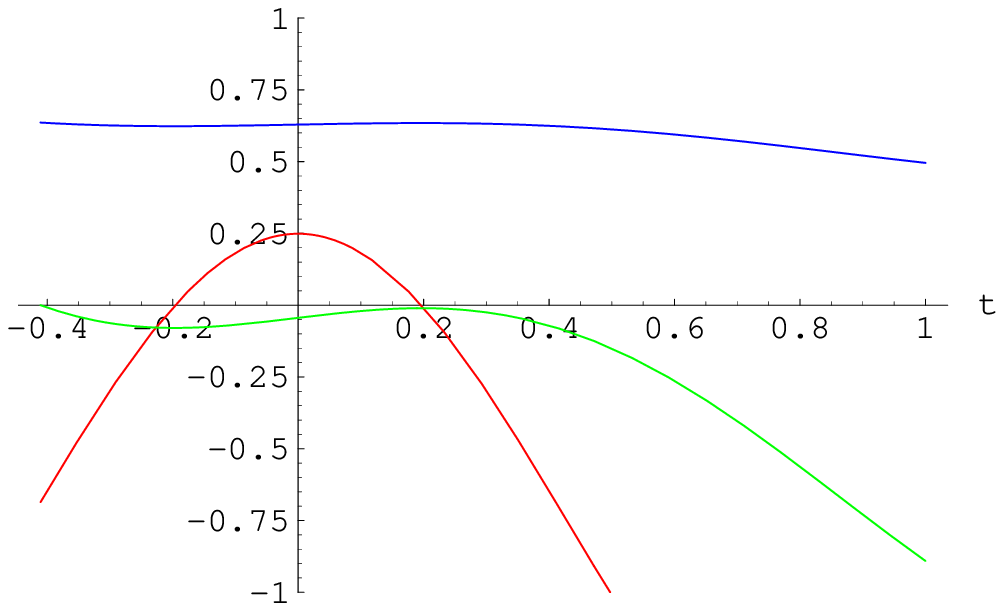}
\inputEps{7cm}{Figure 7: $a=0.01$, $b=0.4$. Red line: $f(t)$. Blue line: $g(t)$.
Green line: $r(t)$}{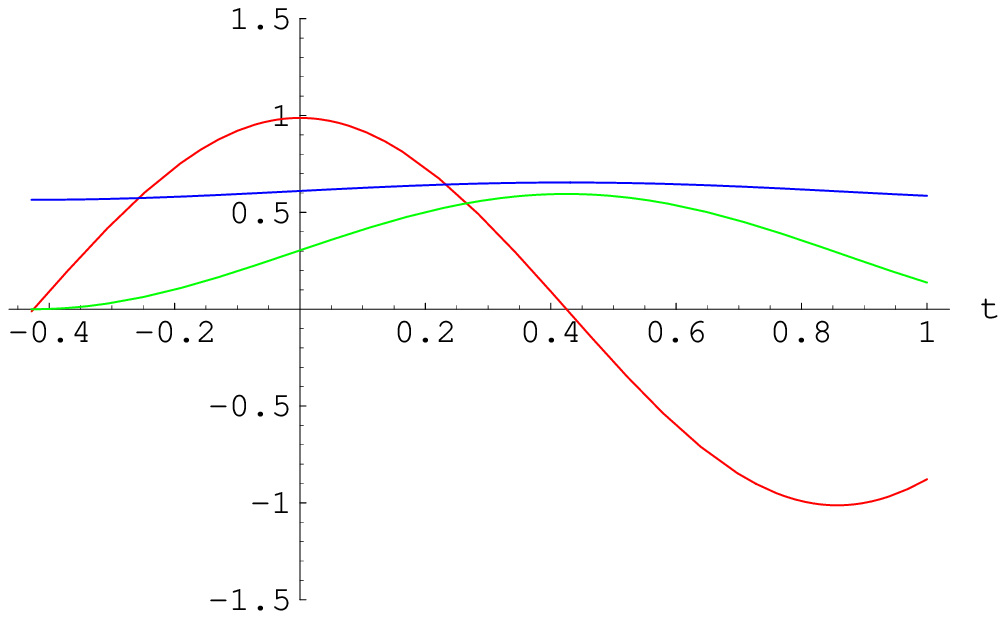}

\bigskip

These results may appear surprising, but can be explained by looking
at the structure of the states $P_\mu(t)$ and their entanglement
characterization. Indeed, one knows that the so-called
\textit{entanglement of formation} \cite{Benn} is a monotonically
increasing function of the concurrence \cite{Woo} and, moreover,
that it cannot increase under local quantum operations. This seems
to conflict with the fact that the local action of
$\gamma_t\otimes{\rm id}$ may increase the concurrence and hence the
entanglement of formation. However, there is no contradiction: in
fact, $\gamma_t$ is not positive, whereas a quantum operation is by
definition completely positive. It is of some interest to inspect in
more detail how non-positivity may generate entanglement by acting
locally. Let us then consider the explicit expression of the
entanglement of formation,
\begin{equation}
\label{ent2} E(\rho)=\min_{\rho=\sum_ip_i\,\rho_i}\sum_ip_i\,
S(\rho^{1}_i)\ ,
\end{equation}
where $S(\rho^1_i)$ is the von Neumann entropy of the marginal
states of party $1$ resulting from partial trace over party $2$,
$\rho^1_i:={\rm Tr}_2(\rho_i)$, obtained from those that contribute
to the convex decomposition $\rho=\sum_ip_i\rho_i$, $p_i\geq 0$,
$\sum_ip_i=1$.

Essentially, $E(\mathbb{G}\otimes{\rm id}[\rho])\leq E(\rho)$ under
a quantum operation $\mathbb{G}$ because any optimal decomposition
$\sum_ip_i\rho_i$ of $\rho$ achieving $E(\rho)$ provides a
decomposition $\sum_ip_i\mathbb{G}\otimes{\rm id}[\rho_i]$ of
$\mathbb{G}\otimes{\rm id}[\rho]$, whence the minimum can only
decrease. In this section we have proved that there are times
$0<s<t$ such that $P_{\mu}(s)=\gamma_s\otimes{\rm id}[P_\mu]$ and
$P_{\mu}(t)=\gamma_t\otimes{\rm id}[P_\mu]$ are well-defined
entangled isotropic states satisfying $E(P_\mu(t))>E(P_\mu(s))$.
From the argument of above, such inequality can hold only if
$\gamma_t\otimes{\rm id}$ acting on an optimal decomposition of
$P_\mu(s)$ does not provide a decomposition of $P_\mu(t)$, namely
only if at least one of the states optimally decomposing $P_\mu(s)$
are not mapped into density matrices by $\gamma_t\otimes{\rm id}$.

To summarize, by resorting to the slippage mechanism, one may force
a non-positive time-evolution $\gamma_t\otimes{\rm id}$ to map a
class of entangled states into states at all times, but it may
happen that this local action increases their entanglement because
it is the positivity of the spectrum of other entangled states,
outside that class, which is spoiled in the course of time.

\section{Conclusion}

We have considered a concrete model of reduced dynamics for a single
qubit weakly interacting with a stochastic environment and derived a reduced
dynamics depending on two parameters.
The resulting semigroup consists of maps $\gamma_t$ that range from
non-positive to positive and completely positive.

The physical meaning of complete positivity is related to the
existence of entangled states and can thus be fully appreciated only
when the open qubit is correlated to an ancilla.
Far from being abstract and out of experimental control, an entangled
qubit pair one of whose parties  experiences a noisy
channel is a typical theoretical and practical context of quantum
information and communication theory.
In these cases, complete positivity cannot be dispensed with,
otherwise physical inconsistencies immediately appear, typically the
presence of negative probabilities in the spectrum of evolving
entangled density matrices which lose their meaning as physical states.

In relation to entanglement, we have studied an approach, the
so-called slippage of initial conditions, whereby complete
positivity is deemed an abstract request devoid of physical content
and as such refused as an unnecessary constraint on the open quantum
dynamics. This means that one prefers to stick to a non-positive
Markovian reduced dynamics and try to cure in some way or the other
its pathological behavior. Indeed, already for a single qubit, one
has to enforce a selection of the admissible initial states which is
supposedly due to the action of the transient regime.

By means of a simple (completely positive) slippage operator
$\mathbb{S}_\mu$, $0\leq\mu\leq 1$, we have turned non-positive
reduced dynamical maps
$\gamma_t$ into completely positive ones,
$\gamma_t\circ\mathbb{S}_\mu$, that eliminate the presence of negative
eigenvalues from the spectrum of time-evolving single qubit states.

However, we have showed that $\gamma_t\otimes{\rm id}$ acting on the
isotropic states $P_\mu=\mathbb{S}_\mu\otimes {\rm id}[P]$, may
increase their entanglement by acting locally. Such an unphysical
possibility is due to the non-positivity of $\gamma_t$ and can be
eliminated by resorting to further slippage operators that may
result in the whole elimination of entangled isotropic states.

Despite the simplicity of the model, we believe it contains the
salient features of a more general structure: sticking to
non-completely positive reduced dynamics, though cured by some
suitable slippage mechanism, would conflict with the presence of
entanglement. In some particular cases discussed in this paper, the
conflict can only be avoided by the drastic elimination, via a
suitable slippage, of whole classes of entangled states.

To conclude, slipped non-(completely) positive semigroups present
two different contradictory aspects in relation to entanglement: on
one hand, the slippage operated by the bath during the transient
phase seems to hamper the possibility of creating entanglement by
embedding two dynamically uncorrelated parties within a same
environment, as discussed in \cite{BF1}. On the other hand, unless
all entangled states are eliminated by the slippage beforehand, the
subsequent Markovian time-evolution may look like being able to
create entanglement by acting locally. However, this entanglement
creation is a spurious artifact: it is due to some entangled state,
not destroyed by the slippage, which develops negative probabilities
while evolving in time. As a consequence, curing non-positive
Markovian evolutions through the slippage operation appears to be
rather unsatisfactory and physically unviable.


\end{document}